\documentclass[showpacs,preprintnumbers,
amsmath,amssymb,aps,prd]{revtex4}
\usepackage{amsfonts}
\usepackage{amsfonts,graphics,epsfig,subfigure}

\begin{document}

\title{Holographic Heat engine within the framework of massive gravity}
\author{Jie-Xiong Mo, Gu-Qiang Li}
 \affiliation{Institute of Theoretical Physics, Lingnan Normal University, Zhanjiang, 524048, Guangdong, China}

\begin{abstract}
Heat engine models are constructed within the framework of massive gravity in this paper. For the four-dimensional charged black holes in massive gravity, it is shown that the heat engines have a higher efficiency for the cases $m^2>0$ than for the case $m=0$ when $c_1<0, c_2<0$. Considering a specific example, we show that the maximum efficiency can reach $0.9219$ while the efficiency for $m=0$ reads $0.5014$. The existence of graviton mass improves the heat engine efficiency significantly. The situation is more complicated for the five-dimensional neutral black holes. Not only the $c_1, c_2, m^2$ exert influence on the efficiency, but also the constant $c_3$ corresponding to the third massive potential contributes to the efficiency. When $c_1<0, c_2<0, c_3<0$, the heat engine efficiency of the cases $m^2>0$ is higher than that of the case $m=0$. By studying the ratio $\eta/\eta_C$, we also probe how the massive gravity influences the behavior of the heat engine efficiency approaching the Carnot efficiency.
\end{abstract}
  \maketitle

\section{Introduction}

      The black hole thermodynamics has gained renewed attention from the perspective of the extended phase space \cite{Dolan2} where the cosmological constant is identified as thermodynamic pressure. Close relation between AdS black holes and van der Waals liquid-gas system has been further enhanced \cite{Kubiznak}. And the mass of the black hole gains new physical interpretation as the enthalpy \cite{Kastor} rather than as the internal energy . For a recent review on the extended phase space thermodynamics, one can refer to Ref. \cite{Kubiznak2}.

      Within the framework of the extended phase space, the creative proposal of holographic heat engine was put forward for the first time in Ref. \cite{Johnson1}, allowing the mechanical work to be extracted from heat energy. And the heat engine cycle can be interpreted holographically as a journal through a family of large $N$ field theories \cite{Johnson1}. This proposal has received considerable attention and the effects of Gauss-Bonnet gravity \cite{Johnson2}, Born-Infeld electrodynamics \cite{Johnson3}, spacetime dimensionality \cite{Belhaj,jiexiong2017,Hennigar} on the heat engine efficiency as well as other interesting aspects \cite{Sadeghi1, Setare, Sadeghi2, Bhamidipati, Johnson4, weishaowen,mengxinhe, Johnson6, zhaoliu} have been probed. Several months ago, Ref. \cite{Johnson5} proposed the idea of approaching the Carnot limit at finite power via the heat engine model constructed in the Reissner-Nordstr\"{o}m (RN)-AdS black holes spacetime. An ingenious heat engine cycle was considered with the critical point placed at one of the corners. It was shown that both the heat engine efficiency $\eta$ and Carnot efficiency $\eta_C$ converge at large $q$. It was further argued that even if the critical point is not on the cycle, the approach $\eta\rightarrow\eta_C$ at large $q$ can be achieved.

    In this paper, we would like to construct the heat engine within the framework of massive gravity. As we know, gravitons are introduced as massless particles in Einstein gravity. However, there exist several arguments that gravitons should be massive. By including mass terms, massive gravity has served as a straightforward modification of general relativity. Fierz and Pauli first constructed the linear theory of massive gravity \cite{Fierz}. It was reported that Boulware-Deser ghost instability~\cite{Boulware1,Boulware2} exists at the nonlinear level. To overcome this problem, Hassan~\cite{Hassan1, Hassan2} proposed the ghost-free massive gravity. Within this framework, black hole solutions with a negative cosmological constant was found~\cite{Vegh}. Their thermodynamics was studied in Refs.~\cite{Caironggen, Adams, huyapeng1} while their $P-V$ criticality was investigated in Ref. \cite{Caoliming}. For the most recent literatures concerning the black holes in massive gravity, see Refs. \cite{Hendi1,Upadhyay,huyapeng2,huyapeng3,huyapeng4,zengxiaoxiong,zoudecheng1,zoudecheng2,panwenjian,Hendi2}, where varieties of peculiar properties were discovered. In this sense, generalizing the current research on heat engine to the framework of massive gravity is of physical significance. It is expected to give rise to new findings of heat engine due to the peculiar properties of massive gravity. Our research may also pave way for improving the heat engine efficiency.

    The organization of this paper is as follows. In Sec.\ref{Sec2} we will review the thermodynamics of black holes in massive gravity. The heat engine model via the four-dimensional black holes in massive gravity will be constructed in Sec.\ref{Sec3} while the five-dimensional black hole case will be discussed in Sec.\ref {Sec4}. In the end, we will present a brief conclusion in Sec.\ref {Sec5}.

\section{A brief review on thermodynamics of black holes in massive gravity}
\label{Sec2}

The action of ($n+2$)-dimensional massive gravity reads~\cite{Vegh}
\begin{equation}
S=\frac{1}{16\pi}\int
d^{n+2}x\sqrt{-g}\Big[R+\frac{n(n+1)}{l^2}-\frac{1}{4}F^2+m^2\sum_{i=1}^4c_i\mathcal{U}_i(g,f)\Big], \label{1}
\end{equation}
where $R$, $F=F_{\mu \nu}F^{\mu \nu }$ and $m$ denote the Ricci scalar curvature, the Maxwell invariant and the massive
parameter respectively. Note that the electromagnetic field tensor $F_{\mu\nu}=\partial_{\mu}A_{\nu}-\partial_{\nu}A_{\mu}$. The main differences between the action of massive gravity and that of Einstein gravity are reflected in the last four massive potential terms associated with the graviton mass. $c_i$ are the constants while $f$ and $\mathcal{U}_i$ denote respectively the reference metric and symmetric polynomials of the eigenvalue of the ($n+2$)$\times$($n+2$) matrix $\mathcal{K}^{\mu}_{~\nu}\equiv\sqrt{g^{\mu\alpha}f_{\alpha\nu}}$. The expressions of $\mathcal{U}_i$ read
\begin{eqnarray}
\mathcal{U}_1&=&[\mathcal{K}],\nonumber\\
\mathcal{U}_2&=&[\mathcal{K}]^2-[\mathcal{K}^2],\nonumber\\
\mathcal{U}_3&=&[\mathcal{K}]^3-3[\mathcal{K}][\mathcal{K}^2]+2[\mathcal{K}^3],\nonumber\\
\mathcal{U}_4&=&[\mathcal{K}]^4-6[\mathcal{K}^2][\mathcal{K}]^2+8[\mathcal{K}^3][\mathcal{K}]+3[\mathcal{K}^2]^2-6[\mathcal{K}^4]. \label{2}
\end{eqnarray}
where $[\mathcal{K}]=\mathcal{K}^{\mu}_{~\mu}$ and the square root in $\mathcal{K}$ can be interpreted as $(\sqrt{A})^{\mu}_{~\nu}(\sqrt{A})^{\nu}_{~\lambda}=A^{\mu}_{~\nu}$. Note that the coefficients $c_i$ might be required to be negative if $m^2>0$ for a self-consistent massive gravity theory~\cite{Caironggen}.

The static black hole solution of the above action was found with the spacetime metric and reference metric read~\cite{Caironggen}
\begin{eqnarray}
ds^2&=&-f(r)dt^2+f^{-1}(r)dr^2+r^2h_{ij}dx^idx^j, \label{3} \\
f_{\mu\nu}&=&\mathrm{diag}(0,0,c_0^2h_{ij}), \label{4}
\end{eqnarray}
where $h_{ij}dx^idx^j$ denotes the line element of an Einstein space with constant curvature $n(n-1)k$.
$k$ can be taken as $1$, $0$, or $-1$ respectively without loss of generality, which corresponds to spherical, Ricci flat, and hyperbolic topology.  $c_0$ is a positive constant. $\mathcal{U}_i$ and $f(r)$ have been obtained as ~\cite{Caironggen}
\begin{eqnarray}
\mathcal{U}_1&=&nc_0/r,\nonumber\\
\mathcal{U}_2&=&n(n-1)c_0^2/r^2,\nonumber\\
\mathcal{U}_3&=&n(n-1)(n-2)c_0^3/r^3,\nonumber\\
\mathcal{U}_4&=&n(n-1)(n-2)(n-3)c_0^4/r^4, \label{5}  \\
f(r)&=&k+\frac{16\pi P}{(n+1)n}r^2-\frac{16\pi
M}{nV_nr^{n-1}}+\frac{(16\pi Q)^2}{2n(n-1)V_n^2r^{2(n-1)}}+\frac{c_0c_1m^2}{n}r+c_0^2c_2m^2\nonumber\\
&~&+\frac{(n-1)c_0^3c_3m^2}{r}+\frac{(n-1)(n-2)c_0^4c_4m^2}{r^2},\label{6}
\end{eqnarray}
where $M$ and $Q$ are the mass and the charge of the black hole respectively while $V_n$ denotes the volume of the space spanned by coordinates $x^i$. In the extended phase space, the pressure $P$ has been identified  as $\frac{n(n+1)}{16\pi l^2}$ and the thermodynamic volume is defined as its conjugate quantity. Solving the equation $f(r_h)=0$ for the largest root, one can express the mass into the function of the horizon radius $r_h$ as~\cite{Caironggen,Caoliming}
\begin{eqnarray}
M&=&\frac{nV_nr_h^{n-1}}{16\pi}\Big[k+\frac{16\pi
P}{(n+1)n}r_h^2+\frac{(16\pi Q)^2}{2n(n-1)V_n^2r_h^{2(n-1)}}+\frac{c_0c_1m^2}{n}r_h+c_0^2c_2m^2\nonumber\\
&~&+\frac{(n-1)c_0^3c_3m^2}{r_h}+\frac{(n-1)(n-2)c_0^4c_4m^2}{r_h^2}\Big].\label{7}
\end{eqnarray}
The Hawking temperature $T$, the entropy $S$, the thermodynamic volume $V$, and the electric potential $\Phi$ have been derived as~\cite{Caironggen,Caoliming}
\begin{eqnarray}
T&=&\frac{1}{4\pi}f'(r_h)=\frac{1}{4\pi r_h}\Big[(n-1)k+\frac{16\pi
P}{n}r_h^2-\frac{(16\pi Q)^2}{2nV_n^2r_h^{2(n-1)}}+c_0c_1m^2r_h+(n-1)c_0^2c_2m^2 \nonumber\\
&~&+\frac{(n-1)(n-2)c_0^3c_3m^2}{r_h}+\frac{(n-1)(n-2)(n-3)c_0^4c_4m^2}{r_h^2}\Big], \label{8}\\
S&=&\int_0^{r_h}T^{-1}\left(\frac{\partial H}{\partial
r}\right)_{Q,P}dr=\frac{V_n}{4}r_h^n,  \label{9}\\
V&=&\left(\frac{\partial H}{\partial
P}\right)_{S,Q}=\frac{V_n}{n+1}r_h^{n+1},  \label{10}\\
\Phi&=&\left(\frac{\partial H}{\partial
Q}\right)_{S,P}=\frac{16\pi}{(n-1)V_nr_h^{n-1}}Q.  \label{11}
\end{eqnarray}

The first law and the Smarr relation of the black hole have been obtained in the extended phase space as~\cite{Caoliming}
\begin{eqnarray}\label{firstlaw}
\mathrm{d}H&=&T\mathrm{d}S+V\mathrm{d}P+\Phi\mathrm{d}Q+\frac{V_nc_0m^2r_h^n}{16\pi}\mathrm{d}c_1+\frac{nV_nc_0^2m^2r_h^{n-1}}{16\pi}\mathrm{d}c_2 \nonumber\\
&~&+\frac{n(n-1)V_nc_0^3m^2r_h^{n-2}}{16\pi}\mathrm{d}c_3+\frac{n(n-1)(n-2)V_nc_0^4m^2r_h^{n-3}}{16\pi}\mathrm{d}c_4,\label{12}\\
(n-1)H&=&nTS-2PV+(n-1)\Phi
Q-\frac{V_nc_0c_1m^2}{16\pi}r_h^n \nonumber\\
&~&+\frac{n(n-1)V_nc_0^3c_3m^2}{16\pi}r_h^{n-2}+\frac{n(n-1)(n-2)V_nc_0^4c_4m^2}{8\pi}r_h^{n-3}. \label{13}
\end{eqnarray}

\section{Four-dimensional charged black holes in massive gravity}
\label{Sec3}

For the four-dimensional black hole (corresponding to $n=2$), $\mathcal{U}_3=\mathcal{U}_4=0$ and $c_3, c_4$ can be set to zero. Then Eq.(\ref{7}) reduces to
\begin{equation}
M=\frac{V_2r_h}{8\pi}\Big[k+\frac{8\pi P}{3}r_h^2+\frac{(8\pi
Q)^2}{V_2^2r_h^2}+\frac{c_0c_1m^2}{2}r_h+c_0^2c_2m^2\Big]. \label{14}
\end{equation}

\begin{figure}
\centerline{\subfigure[]{\label{1a}
\includegraphics[width=8cm,height=6cm]{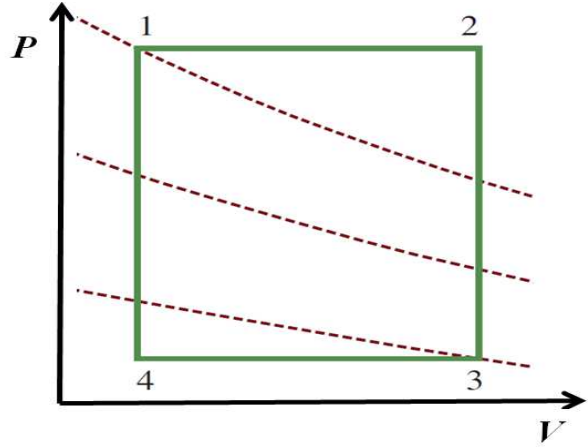}}}
 \caption{The heat engine cycle considered}
\label{fg1}
\end{figure}

Now we define a new kind of heat engine via four-dimensional black holes in massive gravity. As in former literatures \cite{Johnson1, Johnson2, Johnson3}, a rectangle cycle in the $P-V$ plane which consists of two isobars and two isochores is considered here. A sketch picture of the cycle is shown in Fig.\ref{fg1}. Note that the subscripts $1, 2, 3, 4$ denote the relevant physical quantities evaluated at the four corners $1, 2, 3, 4$ respectively.

%%%%%%%%%%%%%%%%%%%%%%%%%%%%%%%%%%%%%%%%%%%%%%%%%%%%%%%%%%%%%%%%%%%%%%%%%%%%%
%%%%%%%%%%%%%%%%%%%%%%%%%%%%%%%%%%%%%%%%%%%%%%%%%%%%%%%%%%%%%%%%%%%%%%%%%%%%%%%%

It can be witnessed from Eqs. (\ref{9}) and (\ref{10}) that the entropy $S$ and the thermodynamic volume $V$ are not independent. It is not difficult to conclude that the specific heat at constant volume $C_V$ equals to zero. Then the isochores are adiabatic.

Via Eqs. (\ref{8}) and (\ref{9}), $T$ can be expressed into the function of $S$ as
 \begin{equation}
T=\frac{-16\pi^2Q^2+32P\pi S^2+kSV_2+c_0m^2(c_0c_2SV_2+2c_1S\sqrt{SV_2})}{8\pi S \sqrt{SV_2}}.\label{15}
\end{equation}%
Then the specific heat at constant pressure $C_P$ can be derived as
 \begin{equation}
C_P=T\left(\frac{\partial S}{\partial T}\right)_{P}=\frac{2S\left[-16\pi^2Q^2+32P\pi S^2+kSV_2+c_0m^2\left(c_0c_2SV_2+2c_1S\sqrt{SV_2}\right)\right]}{48\pi^2Q^2+32P\pi S^2-\left(k+c_0^2c_2m^2\right)SV_2}.\label{16}
\end{equation}%
And the heat input $Q_H$ can be derived through
\begin{equation}
Q_H=\int^{T_2}_{T_1}C_PdT=\int^{S_2}_{S_1}C_P \left(\frac{\partial T}{\partial S}\right)_PdS=\int^{S_2}_{S_1}TdS=\int^{H_2}_{H_1}dH=M_2-M_1,\label{17}
\end{equation}%
where we have adopted the conjecture that the mass $M$ should be interpreted as the enthalpy $H$. The output heat $Q_c$ can be derived in a similar way. Some comments are made in order. The high temperature expansion technique is not necessary here. Both the method utilizing the integral and the method utilizing the exact formula $\eta=1-\frac{M_3-M_4}{M_2-M_1}$ proposed in \cite{Johnson4} can lead to exact result . Both methods are equivalent to each other. However, when one deal with the scheme where $(T_1, T_2, P_1, P_4)$ or $(T_2, T_4, V_2, V_4)$ are specified as operating parameters in the heat engine cycle, only numerical (not analytical) results can be obtained from these two methods. Because in most cases it is also not easy to express the mass $M$ into the function of $T$. One has to solve the equation for numerical results of $S$ or $r_h$ and then substituting them into the expression of $M$.

Utilizing Eqs. (\ref{15}) and (\ref{17}) (or utilizing Eqs. (\ref{9}), (\ref{14}) and (\ref{17}) equivalently), the result of $Q_H$ can be calculated as
\begin{eqnarray}
Q_H&=&\frac{1}{12\pi V_2}\big[3c_0c_1m^2V_2(S_2-S_1)+3(k+c_0^2c_2m^2)V_2^{3/2}(\sqrt{S_2}-\sqrt{S_1}).\nonumber \\
&~&+48\pi^2Q^2\sqrt{V_2}(S_2^{-1/2}-S_1^{-1/2})+32P_1\pi \sqrt{V_2}(S_2^{3/2}-S_1^{3/2})\big].\label{18}
\end{eqnarray}%

The work done along the cycle can be derived as
 \begin{equation}
W=(V_2-V_1)(P_1-P_4)=\frac{8}{3\sqrt{V_2}}(P_1-P_4)\left(S_2^{3/2}-S_1^{3/2}\right).\label{19}
\end{equation}%

Then the heat engine efficiency for four-dimensional charged black holes in massive gravity can be calculated as
 \begin{equation}
\eta=\frac{W}{Q_H}=\left(1-\frac{P_4}{P_1}\right) \times \frac{32P_1\pi (S_1+S_2+\sqrt{S_1S_2})}{B(S_1,S_2,P_1)} ,\label{20}
\end{equation}%
where
 \begin{equation}
B(S_1,S_2,P_1)=32P_1\pi \left(S_1+S_2+\sqrt{S_1S_2}\right)-\frac{48\pi^2Q^2}{\sqrt{S_1S_2}}+3kV_2+3c_0m^2\sqrt{V_2}\left[c_0c_2\sqrt{V_2}+c_1\left(\sqrt{S_2}+\sqrt{S_1}\right)\right].\label{21}
\end{equation}%
It can be examined that this result agrees very well with the result obtained via the exact efficiency formula $\eta=1-\frac{M_3-M_4}{M_2-M_1}$.

From Eq. (\ref{21}), it can be witnessed clearly that the effect of massive gravity is reflected in the fourth term. When $c_1<0, c_2<0$, the heat engine have a higher efficiency for the cases $m^2>0$ than for the case $m=0$. In other words, the graviton mass might help improve the heat engine efficiency.

\begin{figure}
\centerline{\subfigure[]{\label{2a}
\includegraphics[width=8cm,height=6cm]{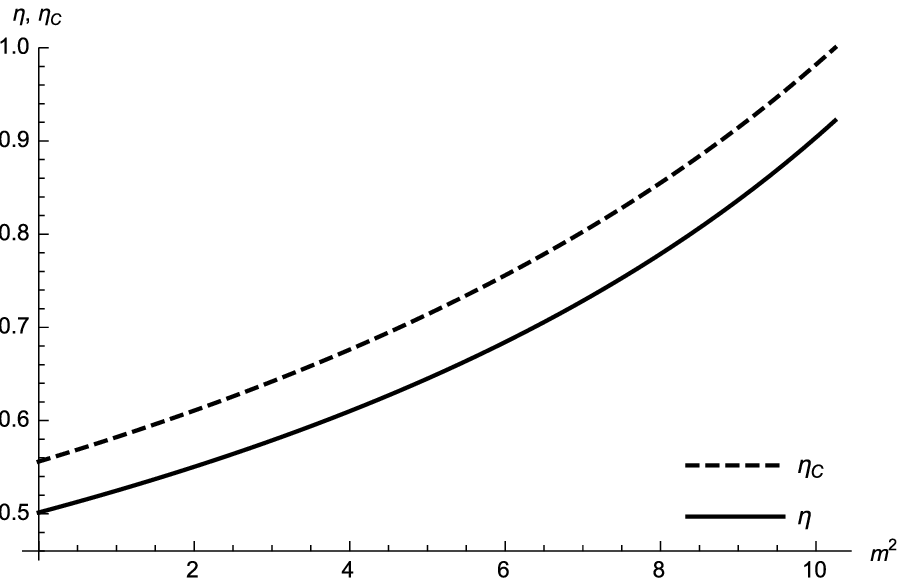}}
\subfigure[]{\label{2b}
\includegraphics[width=8cm,height=6cm]{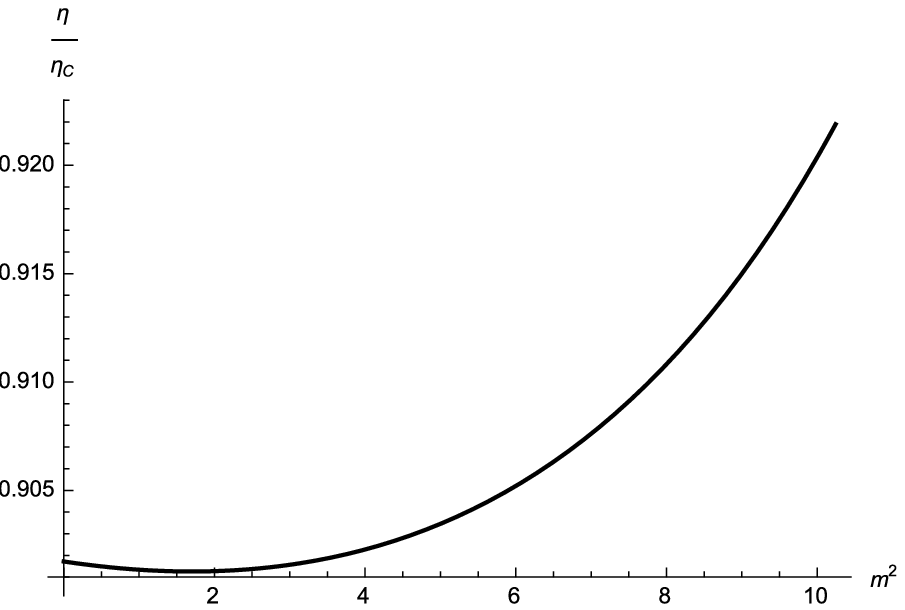}}}
 \caption{A specific example of four-dimensional charged black holes with the parameters chosen as $k=1, c_0=1, c_1=-2, c_2=-3, P_1=2, P_4=1, Q=1, S_1=8, S_2=10$ (a) $\eta, \eta_C$ as the function of $m^2$ (b)$\eta/\eta_C$ as the function of $m^2$}
\label{fg2}
\end{figure}

To sketch a picture of this issue, one can consider a specific example with the parameters chosen as $k=1, c_0=1, c_1=-2, c_2=-3, P_1=2, P_4=1, S_1=8, S_2=10$. To observe the effect of $m^2$ on the heat engine efficiency, we can fix the charge $Q=1$ and let $m^2$ vary from $0$ to $m_0^2$ ($m_0^2$ is the maximum value of $m^2$ that can ensure the positivity of the Hawking temperature). The efficiency $\eta$ as the function of $m^2$ is plotted in Fig.\ref{2a}. It can be witnessed clearly that the efficiency increases with $m^2$. Here, $m_0^2$ can be determined from Eq. (\ref{15}) as $10.2445$. The corresponding maximum efficiency can be obtained as $0.9219$ while the efficiency for $m=0$ takes the value $0.5014$. The existence of graviton mass has improved the heat engine efficiency significantly.

%%%%%%%%%%%%%%%%%%%%%%%%%%%%%%%%%%%%%%%%%%%%%%%%%%%%%%%%%%%%%%%%%%%%%%%%%%%%%
%%%%%%%%%%%%%%%%%%%%%%%%%%%%%%%%%%%%%%%%%%%%%%%%%%%%%%%%%%%%%%%%%%%%%%%%%%%%%%%%

On the other hand, we can fix $m^2=1$ and let $Q$ vary from $0$ to $Q_0$ ($Q_0^2$ is the maximum value of $Q$ that can ensure the positivity of the Hawking temperature) to check the effect of charge when $m^2>0$. From  Fig.\ref{2b}, one can see clearly that the efficiency also increases with $Q$. $Q_0^2$ can be determined from Eq. (\ref{15}) as $6.1187$. The corresponding maximum efficiency can be obtained as $0.8381$ while the efficiency for $Q=0$ takes the value $0.5194$.

To end the discussions here, we would like to remind the reader that one may possibly make a mistake to obtain an efficiency higher than $1$ (obviously not physical) if he does not consider the constraint on the $m^2$ and $Q$ set by the positivity of the Hawking temperature.

\begin{figure}
\centerline{\subfigure[]{\label{3a}
\includegraphics[width=8cm,height=6cm]{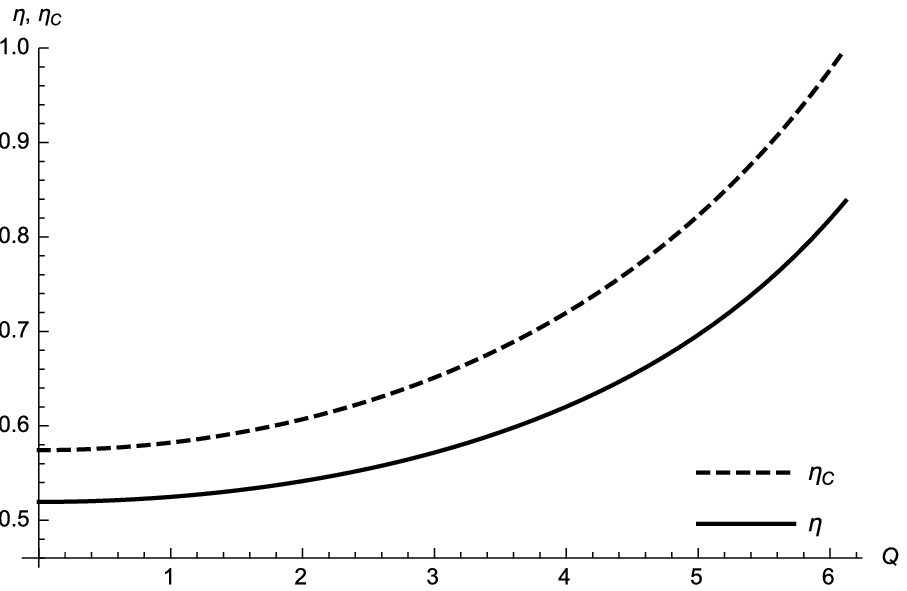}}
\subfigure[]{\label{3b}
\includegraphics[width=8cm,height=6cm]{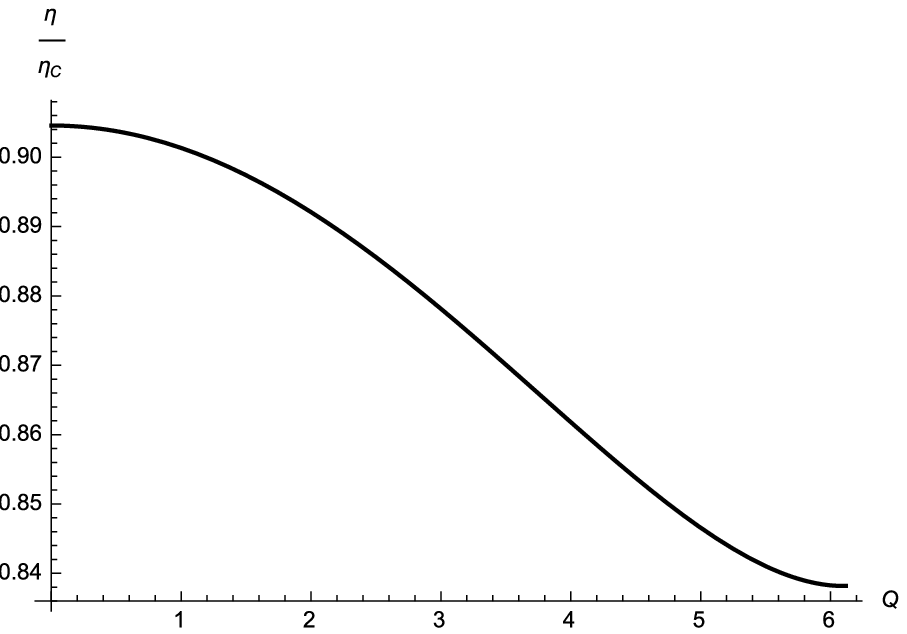}}}
 \caption{A specific example of four-dimensional charged black holes with the parameters chosen as $k=1, c_0=1, c_1=-2, c_2=-3, P_1=2, P_4=1, m^2=1, S_1=8, S_2=10$ (a) $\eta, \eta_C$ as the function of $Q$ (b)$\eta/\eta_C$ as the function of $Q$}
\label{fg3}
\end{figure}

It is also of interest to compare the heat engine efficiency with the well-known Carnot efficiency $\eta_C$, whose formula reads
 \begin{equation}
\eta_C=1-\frac{T_C}{T_H},\label{22}
\end{equation}%
where $T_C, T_H$ denote the lowest and the highest temperature of the heat engine cycle. For the cycle considered in this paper, $T_H=T_2$ and $T_C=T_4$. They can be obtained by substituting $S=S_2, P=P_1$ and $S=S_1, P=P_4$ respectively into the Eq. (\ref{15}). From Figs.\ref{2a} and \ref{3a}, it can be observed that $\eta_C$ increases with $m^2$ and $Q$. When $Q=1$, the maximum $\eta_C$ can approach 1 while the value for $m^2=0$ reads 0.5561.
When $m^2=1$, the maximum $\eta_C$ also approaches 1 while the value for $Q=0$ reads 0.5742. The behavior of $\eta$ approaching $\eta_C$ is depicted in Figs.\ref{2b} and \ref{3b}. With the increasing of $m^2$, the ratio $\eta/\eta_C$ first decreases to a minimum value and then increases monotonically to approach the maximum value 0.9219. However, the effect of $Q$ is quite the reverse. The ratio decreases with $Q$.

%%%%%%%%%%%%%%%%%%%%%%%%%%%%%%%%%%%%%%%%%%%%%%%%%%%%%%%%%%%%%%%%%%%%%%%%%%%%%
%%%%%%%%%%%%%%%%%%%%%%%%%%%%%%%%%%%%%%%%%%%%%%%%%%%%%%%%%%%%%%%%%%%%%%%%%%%%%%%%

\section{Five-dimensional neutral black holes in massive gravity}
\label{Sec4}

For the five-dimensional black hole (corresponding to $n=3$), $\mathcal{U}_4=0$ and $c_4$ can be set to zero. Unlike the four-dimensional case discussed in the former section, $c_3\neq0$. For simplicity, we consider the neutral black holes here (corresponding to $Q=0$). Then Eq.(\ref{7}) becomes
\begin{equation}
M=\frac{3V_3r_h^2}{16\pi}\Big[k+\frac{4\pi
P}{3}r_h^2+\frac{c_0c_1m^2}{3}r_h+c_0^2c_2m^2+\frac{2c_0^3c_3m^2}{r_h}\Big]. \label{23}
\end{equation}

In this section, we would construct the heat engine model via five-dimensional neutral black holes in massive gravity. And a similar rectangle cycle in the $P-V$ plane will be considered.

By the same token, one can quickly get $C_V=0$, implying that no heat flows along the isochores.

Utilizing Eqs. (\ref{8}) and (\ref{9}), the Hawking temperature of five-dimensional neutral black holes can be obtained as
 \begin{equation}
T=\frac{32\times2^{1/3}P\pi S+6k(SV_3^2) ^{1/3}+3c_0m^2\left[2^{1/3}c_0^2c_3V_3+2^{2/3}c_1(S^2V_3) ^{1/3}+2c_0c_2(S V_3^2) ^{1/3}\right]}{12\times2^{2/3}\pi (S^2V_3) ^{1/3}}.\label{24}
\end{equation}%
Then the specific heat at constant pressure $C_P$ can be derived
 \begin{equation}
C_P=T\left(\frac{\partial S}{\partial T}\right)_{P}=\frac{2S\left[-16\pi^2Q^2+32P\pi S^2+kSV_2+c_0m^2\left(c_0c_2SV_2+2c_1S\sqrt{SV_2}\right)\right]}{48\pi^2Q^2+32P\pi S^2-\left(k+c_0^2c_2m^2\right)SV_2}.\label{25}
\end{equation}%

Utilizing Eqs. (\ref{17}) and (\ref{24}) (or utilizing Eqs. (\ref{9}), (\ref{17}) and (\ref{23}) equivalently), the result of $Q_H$ can be calculated as
\begin{eqnarray}
Q_H&=&M_2-M_1=\frac{1}{8\pi }\big[8\times2^{2/3}P_1\pi V_3^{-1/3}(S_2^{4/3}-S_1^{4/3})+3\times2^{1/3}kV_3^{1/3}(S_2^{2/3}-S_1^{2/3})+2c_0c_1m^2(S_2-S_1)\nonumber \\
&~&+3\times2^{2/3}c_0^3c_3m^2V_3^{2/3}(S_2^{1/3}-S_1^{1/3})+3\times2^{1/3}c_0^2c_2m^2V_3^{1/3}(S_2^{2/3}-S_1^{2/3})\big].\label{26}
\end{eqnarray}%

The work $W$ can be obtained as
 \begin{equation}
W=(V_2-V_1)(P_1-P_4)=2^{2/3}V_3^{-1/3}(P_1-P_4)\left(S_2^{4/3}-S_1^{4/3}\right).\label{27}
\end{equation}%

And the heat engine efficiency $\eta$ for five-dimensional neutral black holes can be derived as
 \begin{equation}
\eta=\frac{W}{Q_H}=\left(1-\frac{P_4}{P_1}\right)\times \frac{8\times2^{2/3}P_1\pi \left(S_2^{4/3}-S_1^{4/3}\right)}{C(S_1,S_2,P_1)}  ,\label{28}
\end{equation}%
where
 \begin{eqnarray}
C(S_1,S_2,P_1)&=&c_0V_3^{1/3}m^2\left[2c_1(S_2-S_1)+3\times2^{2/3}c_0^2c_3V_3^{2/3}(S_2^{1/3}-S_1^{1/3})+3\times2^{1/3}c_0c_2V_3^{1/3}(S_2^{2/3}-S_1^{2/3})\right]\nonumber \\
&~&+8\times2^{2/3}P_1\pi \left(S_2^{4/3}-S_1^{4/3}\right)+3\times2^{1/3}kV_3^{2/3}(S_2^{2/3}-S_1^{2/3}).\label{29}
\end{eqnarray}%

From Eq. (\ref{29}), it can be witnessed clearly that the effect of massive gravity is reflected in the first term which contains the factor $m^2$. Obviously, this case is more complicated than the four-dimensional case. Not only the $c_1, c_2, m^2$ exert influence on the efficiency, but also the constant $c_3$ corresponding to the third massive potential contributes to the efficiency. When $c_1<0, c_2<0, c_3<0$, the heat engine efficiency of the cases $m^2>0$ are higher than that of the case $m=0$, just as in the four-dimensional charged black holes discussed in the former section.

\begin{figure}
\centerline{\subfigure[]{\label{4a}
\includegraphics[width=8cm,height=6cm]{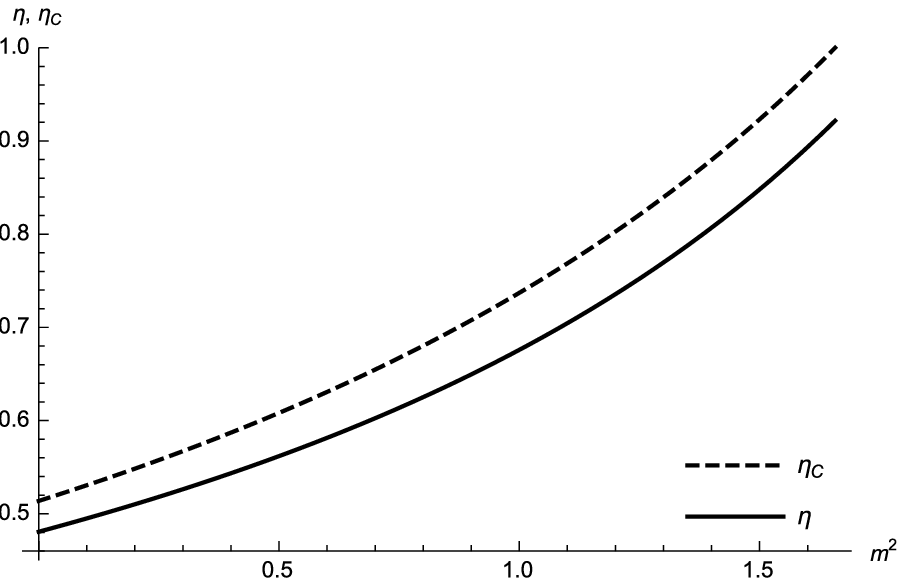}}
\subfigure[]{\label{4b}
\includegraphics[width=8cm,height=6cm]{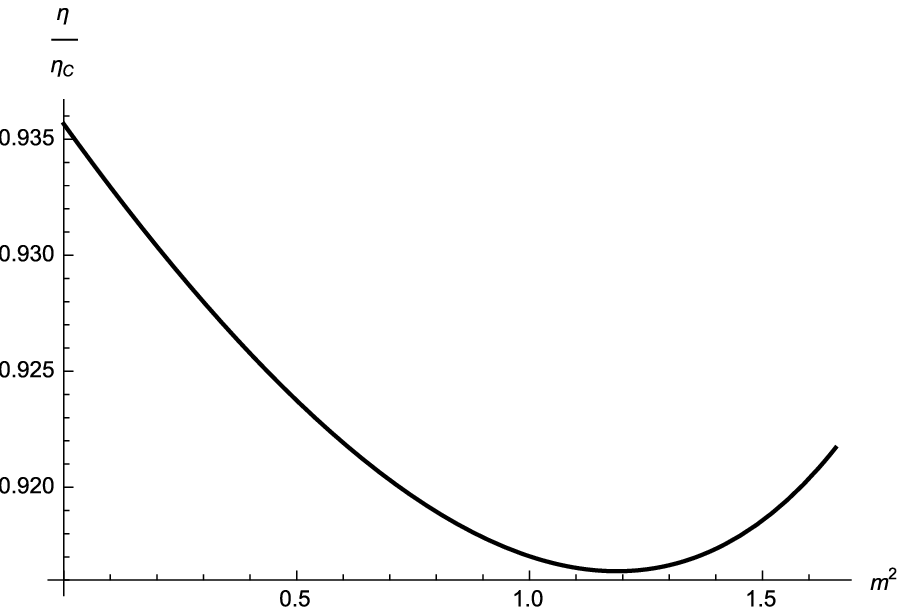}}}
 \caption{A specific example of five-dimensional neutral black holes with the parameters chosen as $k=1, c_0=1, c_1=-2, c_2=-3, c_3=-4, P_1=2, P_4=1, S_1=8, S_2=10$ (a) $\eta, \eta_C$ as the function of $m^2$ (b)$\eta/\eta_C$ as the function of $m^2$}
\label{fg4}
\end{figure}

To gain an intuitive understanding, we present a specific example here with the parameters chosen as $k=1, c_0=1, c_1=-2, c_2=-3, c_3=-4, P_1=2, P_4=1, S_1=8, S_2=10$. To guarantee the positivity of the Hawking temperature, one can first determine the maximum value of $m^2$ (1.6571) via Eq. (\ref{24}). Then one can let $m^2$ vary from $0$ to this maximum value to probe how the efficiency is affected by different choices of $m^2$. The efficiency graph is plotted in Fig.\ref{4a}, from which one can observe that the efficiency increases with $m^2$. The maximum efficiency can be obtained as $0.9217$ comparing to $0.4808$ for $m=0$, showing the impact of the graviton mass on the heat engine efficiency.

One can also make some comparison with Carnot efficiency $\eta_C$, whose behavior is also plotted in Figs.\ref{4a}. As in the four-dimensional charged case, $\eta_C$ increases with $m^2$ while the ratio $\eta/\eta_C$ first decreases monotonically and then increases. However, the maximum value (0.9356) of the ratio $\eta/\eta_C$ is obtained when $m^2=0$, different from four-dimensional charged case.

%%%%%%%%%%%%%%%%%%%%%%%%%%%%%%%%%%%%%%%%%%%%%%%%%%%%%%%%%%%%%%%%%%%%%%%%%%%%%
%%%%%%%%%%%%%%%%%%%%%%%%%%%%%%%%%%%%%%%%%%%%%%%%%%%%%%%%%%%%%%%%%%%%%%%%%%%%%%%%

\section{Conclusions}
\label{Sec5}
    Heat engine models are constructed within the framework of massive gravity. Specifically, we define new kinds of heat engine via four-dimensional charged black holes and five-dimensional neutral black holes in massive gravity. A rectangle cycle in the $P-V$ plane which consists of two isobars and two isochores is considered. The heat input and the work done along the cycle are calculated and the heat engine efficiency is obtained. We argue that both the method utilizing the integral and the method utilizing the exact formula proposed in \cite{Johnson4} can lead to exact result and the high temperature expansion technique is not necessary. Both methods are equivalent to each other. However, when one deal with the scheme where $(T_1, T_2, P_1, P_4)$ or $(T_2, T_4, V_2, V_4)$ are specified as operating parameters in the heat engine cycle, only numerical (not analytical) results can be obtained from these two methods. Because in most cases it is also not easy to express the mass $M$ into the function of $T$. One has to solve the equation for numerical results of $S$ or $r_h$ and then substituting them into the expression of $M$.

    The effect of massive gravity on the heat engine efficiency is probed. For the four-dimensional charged black holes in massive gravity, it is shown that the heat engine have a higher efficiency for the cases $m^2>0$ than for the case $m=0$ when $c_1<0, c_2<0$. In other words, the graviton mass can help improve the heat engine efficiency. A specific example with the parameters chosen as $k=1, c_0=1, c_1=-2, c_2=-3, P_1=2, P_4=1, S_1=8, S_2=10$ is presented. We fix the charge $Q=1$ and let $m^2$ vary from $0$ to $m_0^2$ ($m_0^2$ is the maximum value of $m^2$ that can ensure the positivity of the Hawking temperature). It is shown graphically that the efficiency increases with $m^2$. The maximum efficiency can be obtained as $0.9219$ while the efficiency for $m=0$ takes the value $0.5014$. The existence of graviton mass has improved the heat engine efficiency significantly.

    The case of the five-dimensional neutral black hole is more complicated comparing to the four-dimensional case. Not only the $c_1, c_2, m^2$ exert influence on the efficiency, but also the constant $c_3$ corresponding to the third massive potential contributes to the efficiency. When $c_1<0, c_2<0, c_3<0$, the heat engine efficiency of the cases $m^2>0$ are higher than that of the case $m=0$. We present an example with the parameters chosen as $k=1, c_0=1, c_1=-2, c_2=-3, c_3=-4, P_1=2, P_4=1, S_1=8, S_2=10$. The maximum efficiency can be obtained as $0.9217$ comparing to $0.4808$ for $m=0$, showing the impact of the graviton mass on the heat engine efficiency.

    It is also of interest to compare the heat engine efficiency with the Carnot efficiency. For the four-dimensional black holes, with the increasing of $m^2$, the ratio $\eta/\eta_C$ first decreases to a minimum value and then increases monotonically to approach the maximum value 0.9219. However, the effect of $Q$ is quite the reverse. The ratio decreases with $Q$. This peculiar property may be attributed to the combined effect of the charge and the massive gravity. The qualitative behavior of $\eta/\eta_C$ is similar for the five-dimensional neutral black hole case. However, the maximum value of the ratio $\eta/\eta_C$ is obtained when $m^2=0$, different from four-dimensional case.

 \section*{Acknowledgements}

 This research is supported by National Natural Science Foundation of China (Grant No.11605082), and in part supported by Natural Science Foundation of Guangdong Province, China (Grant Nos.2016A030310363, 2016A030307051, 2015A030313789).

\end{document}